\documentclass[12pt,draftcls,onecolumn,print]{ieeecolor}
\usepackage{generic}
\usepackage{cite}
\usepackage{amsmath,amssymb,amsfonts,amsbsy}
\usepackage{algorithmic}
\usepackage{graphicx}
\usepackage{hyperref}
\hypersetup{hidelinks=true}
\usepackage{cleveref}

\newtheorem{theorem}{Theorem}

\newtheorem{proposition}{Proposition}

\newtheorem{corollary}{Corollary}

\newtheorem{definition}{Definition}

\newtheorem{assumption}{Assumption}

\newtheorem{lemma}{Lemma}

\newtheorem{remark}{Remark}

\newtheorem{example}{Example}

\crefname{assumption}{assumption}{assumptions}
\crefformat{equation}{(#2#1#3)}
\Crefname{equation}{}{}

\usepackage[skip=0pt]{caption}
\usepackage{subcaption}
\usepackage{diagbox}
\def\R{{\mathbb{R}}}   

\def\E{{\mathbb{E}}}

\def\1{{\mathbf{1}}}

\newcommand{\inner}[2]{\langle #1, #2 \rangle}
\DeclareMathOperator*{\argmin}{arg\,min}

\DeclareMathOperator*{\cvar}{CVaR}

\newcommand*{\QEDB}{\null\nobreak\hfill\ensuremath{\square}}%

\usepackage{textcomp}
\def\BibTeX{{\rm B\kern-.05em{\sc i\kern-.025em b}\kern-.08em
    T\kern-.1667em\lower.7ex\hbox{E}\kern-.125emX}}
\begin{document}
\title{On Structural Properties of Risk-Averse Optimal Stopping Problems}
\author{Xingyu Ren, Michael C. Fu, and Steven I. Marcus
\thanks{Xingyu Ren and Steven I. Marcus are with the Department of Electrical and Computer Engineering and Institute for Systems Research, University of Maryland, College Park, MD 20742, USA (e-mail: \{renxy,marcus\}@umd.edu). }
\thanks{Michael C. Fu is with the Robert H. Smith School of Business and Institute for Systems Research, University of Maryland, College Park, MD 20742, USA (e-mail: mfu@umd.edu).}}


\maketitle

\begin{abstract}
    We establish structural properties of optimal stopping problems under time-consistent dynamic (coherent) risk measures, focusing on value function monotonicity and the existence of control limit (threshold) optimal policies. While such results are well developed for risk-neutral (expected-value) models, they remain underexplored in risk-averse settings. Coherent risk measures (e.g., conditional value-at-risk (CVaR), mean–semideviation) typically lack the tower property and are subadditive rather than additive, complicating structural analysis. We show that value function monotonicity mirrors the risk-neutral case. Moreover, if the risk envelope associated with each coherent risk measure admits a minimal element, the risk-averse optimal stopping problem reduces to an equivalent risk-neutral formulation. We also develop a general procedure for identifying control limit optimal policies and use it to derive practical, verifiable conditions on the risk measures and MDP structure that guarantee their existence. We illustrate the theory and verify these conditions through optimal stopping problems arising in operations, marketing, and finance.
\end{abstract}

\begin{IEEEkeywords}
Markov decision processes, optimal stopping, coherent risk measures, structural properties, control limit policy
\end{IEEEkeywords}

\section{Introduction}

Optimal stopping problems are a fundamental class of sequential decision-making models in which a decision-maker chooses when to terminate a stochastic process to maximize cumulative reward or minimize cumulative cost. Classical formulations assume risk neutrality, optimizing only the expectation of the cumulative payoff. In high-stakes domains such as finance, insurance, healthcare, and supply chain management, decision-makers are often risk-averse, seeking to limit volatility and prevent consequential losses from rare events. Incorporating risk measures into these models is therefore essential.

Coherent risk measures provide a rigorous framework for quantifying and managing risk, satisfying practically interpretable properties such as monotonicity, convexity, and positive homogeneity \cite{artzner1999coherent}. Moreover, they admit dual representations, under which risk-minimization problems can be formulated as distributionally robust stochastic optimization problems \cite{ruszczynski2006optimization}. Subsequently, \cite{ruszczynski2010risk} introduced finite-horizon risk-averse Markov decision processes (MDPs) with time-consistent dynamic risk measures. In particular, when dynamic risk measures are defined as compositions of conditional coherent risk measures \cite{ruszczynski2006conditional}, both time consistency and Bellman equations hold, enabling risk-averse dynamic programming (DP). This framework was further generalized to transient models \cite{cavus2014risk} and to optimal stopping problems \cite{pichler2022risk}.

Structural properties are \textit{a priori} characterizations of the optimal policy or value function in an MDP---such as monotone value functions and control limit (threshold) optimal policies---that can be obtained without computing them exactly, a task often computationally burdensome for large-scale problems. In risk-neutral settings, structural properties are well documented across applications such as inventory management, organ transplantation, option pricing, and maintenance \cite{veinott1966opimality,ren2024optimal,wu2003optimal,drent2024condition}. Structural properties provide qualitative insight and can be embedded in algorithms to accelerate computation and convergence while reducing sample/data requirements, for example in approximate dynamic programming (ADP) \cite{jiang2015approximate}, stochastic approximation (SA) \cite{wu2003optimal}, and linear programming (LP) \cite{mattila2017computing}.


In risk-neutral MDPs, frameworks for establishing structural properties are well developed. Regarding value functions, \cite{smith2002structural} showed that when the transition law and one-step cost function both satisfy a common ``closed convex cone'' (C3) property---examples include monotonicity, modularity, and convexity---the value function inherits the corresponding property. For solution structure in general parameterized optimization problems, \cite{topkis1978minimizing} established that joint modularity in the parameter and optimization variables guarantees the existence of monotone optimal solutions, laying the foundation for subsequent work on monotone policies (with control limit policies as a special case) in risk-neutral MDPs \cite{puterman2014markov,serfozo1976monotone}, partially observable MDPs \cite{lovejoy1987some,miehling2020monotonicity}, and risk-sensitive MDPs with exponential utility \cite{brau1997controlled}. For risk-averse MDPs within the framework of \cite{ruszczynski2010risk}, structural results have been derived for certain inventory models \cite{ahmed2007coherent,yang2017monotone}, but a general methodology, even for optimal stopping problems, remains lacking.

In this paper, we address this gap for finite-horizon risk-averse optimal stopping under time-consistent dynamic risk measures \cite{ruszczynski2010risk}, focusing on value function monotonicity and the existence of control limit optimal policies. Compared with general MDPs, optimal stopping has a distinct control and cost structure: only two actions (stop or continue), with continuation yielding uncontrolled Markovian evolution. In applications, the continuation cost usually reflects only short-term (e.g., one-period) effects, whereas the terminal cost aggregates long-run effects and can be orders of magnitude larger (e.g., \cite{ren2024optimal}); in some financial models \cite{wu2003optimal}, only a terminal cost is present. In \Cref{sec:control-limit}, we show that these distinctions make frequently used conditions for identifying structure in general MDPs inapplicable. Although existing studies \cite{oh2016characterizing} provide general guidelines tailored to risk-neutral optimal stopping, extending these results to the risk-averse setting is nontrivial because (i) expectation is additive whereas coherent risk measures are only subadditive, and (ii) (conditional) risk measures do not satisfy the tower property---so risk-neutral arguments based on additivity or iterated conditioning break down. Even so, DP remains valid in the risk-averse framework \cite{ruszczynski2010risk}, and thus backward induction---the central technique for establishing structural results---is still applicable, making analogous structural results achievable with assumptions and proofs refined for the risk-averse framework. 
We summarize our main contributions as follows:
\begin{itemize}
\item \textbf{Value function monotonicity.} We show that classical risk-neutral arguments extend to risk-averse optimal stopping under mild regularity of the risk measures: if the transition law and one-step cost functions satisfy suitable monotonicity assumptions, then the value function is monotone \cite{smith2002structural}---either \textit{jointly} across all state dimensions or \textit{componentwise} along selected dimensions, depending on the assumption. For the joint case, when the risk envelope of each (conditional) coherent risk measure admits a minimal element, the risk-averse problem reduces to an equivalent risk-neutral formulation. For the componentwise case, standard risk-neutral arguments \cite{oh2016characterizing,jiang2015approximate} that rely on the tower property fail; we provide an alternative coupling-based proof.
\item \textbf{Existence of control limit optimal policies.} The conventional risk-neutral framework \cite{oh2016characterizing} does not apply because coherent risk measures are only subadditive. We therefore develop a modified framework---tailored to coherent risk measures and compatible with subadditivity---to identify when control limit optimal policies exist. Within this framework, we derive verifiable sufficient conditions in two settings: (i) both the (conditional) coherent risk measures and the state vectors satisfy suitable comonotonicity conditions, and (ii) a one-step look-ahead policy is optimal.
\end{itemize}

The remainder of the paper is organized as follows. \Cref{sec:formulation} formalizes the risk-averse optimal stopping problem and reviews preliminaries on risk measures. \Cref{sec:monotone} establishes conditions under which the value function is monotone---jointly or componentwise (along selected dimensions). \Cref{sec:control-limit} develops a general verification framework for control limit optimal policies, and, building on it, derives practical, verifiable sufficient conditions. \Cref{sec:conc} concludes. Throughout \Cref{sec:monotone,sec:control-limit}, we include illustrative examples from operations, marketing, and finance.

\section{Risk-averse Optimal Stopping Problem}\label{sec:formulation}
In this section we (i) formally define the optimal stopping problem within the MDP framework, (ii) review time-consistent dynamic risk measures and coherent risk measures, and (iii) present simple examples illustrating subadditivity and the failure of the tower property for coherent risk measures---features that preclude a straightforward carryover of structural results from the risk-neutral setting.

\subsection{Optimal Stopping Problem Formulation}
Consider a finite time-horizon $\{0,\dots,T\}$ and a Markov process $\{X_t\}_{t=0}^T$ on a probability space $(\Omega,\mathcal F, P_0)$ with reference measure $P_0$. Let $\mathcal X\cup \{\mathcal T\}$ be the state space, where $\mathcal X\subseteq \R^n$ and $\mathcal T$ is an absorbing terminal state (no further cost is incurred there). Let $\mathcal P(\mathcal X)$ denote the set of probability measures on $\mathcal X$. At each $t\in\{0,\cdots,T-1\}$, the decision-maker chooses an action $u_t\in\mathcal A=\{S,C\}$ (the action space), where $S$ (stop) terminates the process and moves the state to $\mathcal T$, and $C$ (continue) advances the process according to a (possibly time-dependent) Markov transition kernel $Q_t(\cdot|x_t)\in \mathcal P(\mathcal X)$ that depends on the current state $X_t=x_t\in\mathcal X$. Let $\{\mathcal F_t\}_{t=0}^T$ be the filtration generated by $\{X_t\}_{t=0}^T$. Given $X_t=x_t\in\mathcal X$, if stopping at time $t$, a terminal cost $s_t(x_t)$ is incurred; otherwise, a continuation cost $c_t(x_t)$ is incurred, where $s_t,c_t:\mathcal X\mapsto\R$. Define the one-period cost $z_t(x,u):=s_t(x)\1\{u=S\}+c_t(x)\1\{u=C\}$. We consider deterministic Markov policies $\mathcal D:=\{d=(d_0,\cdots,d_{T-1}) \ | \ d_t:\mathcal X\mapsto \mathcal A\}$. For a policy $d\in\mathcal D$, define the stopping time $\tau_d:=\min\{t\leq T \ | \ d_t(X_t)=S \}$ adapted to $\{\mathcal F_t\}_{t=0}^T$. Let $\mathcal Z_t$ denote the space of $\mathcal F_t$-measurable random variables and $Z_t:=z_t(X_t,d(X_t))\1\{t\leq\tau\}\in \mathcal Z_t$ as the period-$t$ cost. The total cost is $\sum_{t=1}^T Z_t=s_\tau(X_\tau)+\sum_{t=0}^{\tau-1}c_t(X_t)$.

\subsection{Risk-Averse Optimal Stopping}\label{sec:coherent-def}
To evaluate risk of the cost sequence $\{Z_t\}_{t=0}^T$, we adopt the time-consistent dynamic risk measures proposed in \cite{ruszczynski2010risk}. Let $\mathcal Z_{t,T}=\mathcal Z_t\times\cdots\times\mathcal Z_T$. A \textit{dynamic risk measure} is a sequence of conditional mappings $\rho_{t,T}:\mathcal Z_{t,T}\mapsto\mathcal Z_t$ for $t=0,\dots,T-1$ that evaluates the risk of the future cost stream $(Z_t,\dots,Z_T)$ from the perspective of time $t$. A key property is \textit{time consistency}: if two cost streams coincide up to some time and one is deemed riskier thereafter, then this risk ordering must already hold at the current time. Under time consistency and suitable translation properties, \cite{ruszczynski2010risk} shows that $\rho_{t,T}$ admits the following nested form:
\begin{align*}
    \rho_{t,T}\left(Z_t,\cdots,Z_T\right)
    &=Z_t+\rho_t(Z_{t+1}+\rho_{t+1}(Z_{{t+2}}+\cdots\\
    &+\rho_{T-2}(Z_{T-1}+\rho_{T-1}(Z_T))\cdots)),
\end{align*}
where the one-step mapping $\rho_t:\mathcal Z_{t+1}\mapsto\mathcal Z_t$ is given by $\rho_t(\cdot):= \rho_{t,t+1}(0,\cdot)$. If, in addition, each $\rho_t$ satisfies the coherence axioms (see \Cref{def:coherent}), then the dynamic risk measure admits the nested compositional form:
\begin{align}\label{eq:risk-measure-2}
    \rho_{t,T}\left(Z_t,\cdots,Z_T\right)=\rho_t\circ\cdots\circ\rho_{T-1}\left(\sum_{\tau=t}^{T}Z_t\right).
\end{align}
In the remainder of this paper, we focus on dynamic risk measures of the above form with each $\rho_t$ coherent. Although this is a subclass rather than the most general case, it yields DP equations, a foundational tool for developing structural results. For a fixed initial state $X_0=x_0\in\mathcal X$ and policy $d\in\mathcal D$, the risk of the corresponding cost sequence $\{Z_t\}_{t=0}^T$ is
\begin{align*}
    &J(d,x_0):=z_0(x_0,d_0(x_0))\\
    &+\rho_0(z_1(X_1,d_1(X_1))+\rho_1(z_2(X_2,d_2(X_2))\\
    &+\cdots+z_{T-1}(X_{T-1},d_{t-1}(X_{T-1}))+\rho_{T-1}(s_T(X_T))\cdots)).
\end{align*}
For fixed $x\in\mathcal X$, let $\tilde X_{t+1}(x)$ denote a random variable with law $Q_t(\cdot| x)$, denoted by $\tilde X_{t+1}(x)\sim Q_t(\cdot | x)$, representing the transition from state $x$ at time $t$. Under suitable regularity assumptions on the transition kernels, one-step risk mappings, and cost functions \cite{ruszczynski2010risk}, the risk minimization problem $v_0(x):=\min_{d\in\mathcal D} J(d,x)$ is solved by the DP recursion
\begin{align}\label{eq:dp}
    \begin{split}
        &v_t(x)=\min\{s_t(x),c_t(x)+\rho_t(v_{t+1}(\tilde X_{t+1}(x))) \},~t<T,\\
        &v_T(x)=s_T(x).
    \end{split}
\end{align}
We call $\{v_t\}_{t=0}^T$ the sequence of \textit{value functions} and any policy $d^*\in\argmin_{d\in\mathcal D} J(d,x),~\forall x\in\mathcal X$ an \textit{optimal policy}.

\subsection{Dual Representation of Coherent Risk Measures}
\begin{definition}\label{def:coherent}
    A one-step (conditional) risk measure $\rho_t:\mathcal Z_{t+1}\mapsto\mathcal Z_t$ is \textit{coherent} if it satisfies:
    \begin{itemize}
    \item Convexity: $\rho_t(\lambda Z + (1-\lambda)W )\leq \lambda \rho_t (Z) + (1-\lambda)\rho_t (W)$, $\forall\lambda\in[0,1],~Z,W\in \mathcal Z_{t+1}$.
    \item Monotonicity: if $Z\leq W$ almost surely (a.s.), then $\rho_t(Z)\leq \rho_t(W)$, $\forall Z,W\in \mathcal Z_{t+1}$.
    \item Translational invariance: $\rho_t(Z+W)=Z+\rho_t(W)$, $\forall Z\in \mathcal Z_{t}, ~W\in \mathcal Z_{t+1}$.
    \item Positive homogeneity: $\rho_t(\lambda W)=\lambda \rho_t(W),~\forall W\in \mathcal Z_{t+1},~\lambda\geq 0$.
    \end{itemize}
\end{definition}
Let $v:\mathcal X\mapsto \R$ be measurable. Fix $t$ and $x_t\in\mathcal X$. If $\rho_t$ is coherent, then $\rho_t\big(v\big(\tilde X_{t+1}(x_t)\big)\big)$ admits the following dual representation \cite{ruszczynski2010risk}:
\begin{align}\label{eq:dual-rep}
    \rho_t\big(v\big(\tilde X_{t+1}(x_t)\big)\big)= \sup_{P\in \mathcal A_t(Q_t(\cdot | x_t))} \inner{v}{P},
\end{align}
where $\mathcal A_t(Q_t(\cdot | x_t))\subset \mathcal P (\mathcal X) $ is the \textit{risk envelope}---a closed, convex set of probability measures associated with $\rho_t$ and the transition law $Q_t(\cdot | x_t)$, and $\inner{v}{P}:=\int_\mathcal{X} v(x)P(dx)$. This dual representation interprets a coherent risk measure as a ``worst-case expectation'' over the risk envelope, thereby linking risk-averse MDPs to distributionally robust MDPs \cite{wiesemann2013robust}. Conversely, any closed, convex subset of $\mathcal P (\mathcal X)$ defines a coherent risk measure via the dual representation. The following example illustrates the dual representation of conditional value-at-risk (CVaR), one of the most popular coherent risk measures. See \cite{ruszczynski2010risk} for additional examples.
\begin{example}\label{exp:cvar-dual}
    For $W\in\mathcal Z_{t+1}$ and tail level $\alpha\in(0,1]$, the conditional CVaR at time $t$, denoted by $\cvar_{\alpha,t}(W)$, is the $\mathcal F_t$-conditional mean of $W$ over its worst $\alpha$-tail. $\cvar_{\alpha,t}$ is coherent and admits the Rockafellar–Uryasev representation \cite{rockafellar2000optimization}: $\cvar_{\alpha,t}(W)=\inf_{U\in\mathcal Z_t} \{U+\E((W-U)_+|\mathcal F_t)/\alpha\}$. In \Cref{eq:dual-rep}, with $\rho_t:=\cvar_{\alpha,t}$, the associated risk envelope is $\mathcal A_t(Q_t(\cdot | x_t))=\{P\in\mathcal \mathcal P(\mathcal X) \ | \ P \ll Q_t(\cdot | x_t), ~ dP/dQ_t(\cdot | x_t) \leq1/\alpha\}$, where $dP/dQ_t(\cdot | x_t)$ is the Radon–Nikodym derivative (the likelihood ratio---the ratio of probability mass functions (pmfs) in the discrete case or of densities in the continuous case), and $P\ll Q_t(\cdot | x_t)$ denotes that $P$ is absolutely continuous with respect to (w.r.t.) $Q_t(\cdot | x_t)$.\QEDB
\end{example}

\subsection{Breakdown of Tower Property and Additivity for Coherent Risk Measures}
We present two counterexamples showing that coherent risk measures may lack two basic properties of expectation---the tower property and additivity. Throughout, we use $\rho_t=\cvar_{\alpha,t}$ as a running example, writing $\cvar_{\alpha}(\cdot|\mathcal F_t)$ in place of $\cvar_{\alpha,t}(\cdot)$. For $Z\in\mathcal Z_{t+2}$, whereas $\E(\E(Z|\mathcal F_{t+1})|\mathcal F_t)=\E(Z|\mathcal F_t)$, the example below shows that, in general, $\cvar_{\alpha}(\cvar_{\alpha}(Z|\mathcal F_{t+1})|\mathcal F_t)\neq\cvar_{\alpha}(Z|\mathcal F_t)$.
\begin{example}
    Let $\alpha=0.05$ and define the pmf of $(X_1,X_2)$ by
    \begin{align*}
        p(x_1,x_2)=\begin{cases}
            0.03, &(x_1,x_2)=(-80,100),\\
            0.02, &(x_1,x_2)=(0,100),\\
            0.95, &(x_1,x_2)=(0,0).
        \end{cases}
    \end{align*}
    Then, $\cvar_{\alpha}(X_1+X_2|\mathcal F_0)=(20\times0.03+100\times 0.02)/0.05=52$. Conditioning on $X_1$, we have $\cvar_{\alpha}(X_2|X_1=-80)=100$ and $\cvar_{\alpha}(X_2|X_1=0)=100\times\frac{0.02}{0.02+0.95}\times\frac{1}{0.05}=41.24$. Hence $\cvar_\alpha(X_1+X_2| X_1)=X_1+\cvar_\alpha(X_2|X_1)$ takes $20$ with probability (w.p.) $0.03$ and $41.24$ w.p. $0.97$, and thus $\cvar_\alpha(\cvar_\alpha(X_1+X_2| \mathcal F_1)|\mathcal F_0)=41.24\neq \cvar_{\alpha}(X_1+X_2|\mathcal F_0)$. \QEDB
\end{example}

The next example shows that CVaR can be strictly subadditive.
\begin{example}
    Let $\alpha=0.05$ and define the pmf of $(X_1,X_2)$ by
    \begin{align*}
        p(x_1,x_2)=\begin{cases}
            0.05, &(x_1,x_2)=(100,0),\\
            0.05, &(x_1,x_2)=(0,100),\\
            0.90, &(x_1,x_2)=(0,0).
        \end{cases}
    \end{align*}
    Then, $\cvar_\alpha(X_1|\mathcal F_0)=\cvar_\alpha(X_2|\mathcal F_0)=\cvar_\alpha(X_1+X_2|\mathcal F_0)=100$. Therefore, $\cvar_\alpha(X_1+X_2|\mathcal F_0)< \cvar_\alpha(X_1|\mathcal F_0)+\cvar_\alpha(X_2|\mathcal F_0)$. \QEDB
\end{example}

The absence of the tower property and additivity makes some arguments commonly used to establish structural results in risk-neutral settings inapplicable. Without the tower property, one cannot ``analyze the target dimension by conditioning on (fixing) the rest,'' a standard step for proving structure in a designated dimension (e.g., \cite[Proposition 3]{oh2016characterizing} and \cite[Proposition 1]{jiang2015approximate}). Likewise, nonadditivity makes it impossible to apply horizon-wise backward induction to propagate (super/sub)modularity of Q-functions---arguments frequently employed to prove monotone optimal policies \cite{serfozo1976monotone}. We will revisit these distinctions when proving the specific structural results. In \Cref{sec:monotone,sec:control-limit}, we show that analogous structural results remain attainable, but require assumptions and/or arguments tailored for risk-averse settings.

\section{Monotonicity of the value function}\label{sec:monotone}
In this section, we establish conditions under which the value functions are monotone either across all state dimensions or componentwise on a designated subset. We begin by formally defining stochastic monotonicity on a partially ordered space, a key building block.
\begin{definition}[Stochastic ordering and monotonicity]\label{def:stochastic-mono}
    Let $\mathcal X\subseteq\R^n$ be equipped with a partial order $\preceq$.
    \begin{itemize}
        \item A function $v:\mathcal X\mapsto\R$ is \textit{increasing w.r.t. $\preceq$} if $x\preceq x'\implies v(x)\leq v(x')$ (resp., \textit{decreasing} if $v(x)\geq v(x')$). 
        \item For $\mathcal X$-valued random variables $X,Y$, we say that $X$ \textit{first-order stochastically dominates} (FOSD) $Y$, denoted by $X\succeq_{sd}Y$, if $\E(v(X))\geq \E(v(Y))$ for every increasing $v:\mathcal X\mapsto\R$.
        \item For $x\in\mathcal X$, write $\tilde X(x)\sim Q(\cdot| x)$. A transition kernel $Q(\cdot| x)\in\mathcal P(\mathcal X)$ is \textit{stochastically increasing} (resp., decreasing) if $x\preceq x'\implies\tilde X(x)\preceq_{sd}\tilde X(x')$ (resp., $\tilde X(x)\succeq_{sd}\tilde X(x')$). Equivalently, we write $Q(\cdot| x)\preceq_{sd}Q(\cdot| x')$ (with a slight abuse of notation).    
    \end{itemize}
\end{definition}
The definition above generalizes the classical notion of FOSD on a totally ordered space \cite{ross1996stochastic}: for real-valued random variables $X_1,X_2$, we have $X_1 \succeq_{sd} X_2$ if and only if $F_1(x)\leq F_2(x)$, or equivalently, $\overline F_1(x)\geq \overline F_2(x)$, $\forall x\in\R$, where $F_i$ is the cumulative distribution function (CDF) of $X_i$ and $\overline F_i:=1-F_i$. Stochastic monotonicity captures that a smaller current state shifts the conditional next-state distribution toward smaller states and is widely used, e.g., in reliability engineering. As with expectation, coherent risk measures preserve stochastic ordering when they are \textit{distribution-invariant} (introduced below); commonly used coherent measures---e.g., CVaR, entropic value-at-risk (EVaR), and mean–semideviation---are distribution-invariant.
\begin{definition}
   A one-step risk mapping $\rho_t:\mathcal Z_{t+1}\mapsto\mathcal Z_t$ is \textit{distribution-invariant} if $Z\stackrel{d}{=}W$ implies $\rho_t(Z)=\rho_t(W),~\forall Z,W\in\mathcal Z_{t+1}$, where $\stackrel{d}{=}$ denotes ``equal in distribution''.
\end{definition}
The following result, proved in \cite{ruszczynski2006optimization}, implies that distribution-invariant coherent risk measures preserve stochastic ordering.
\begin{lemma}\label{lemma:trans-inv}
    Suppose $\rho_t:\mathcal Z_{t+1}\mapsto\mathcal Z_t$ is distribution-invariant. Then $\rho_t$ is consistent with FOSD, i.e., $Z\preceq_{sd} W\implies\rho_t(Z)\leq \rho_t(W),~\forall Z,W\in\mathcal Z_{t+1}$, if and only if it satisfies the monotonicity axiom in \Cref{def:coherent}.
\end{lemma}
For the remainder of the paper, we assume the one-step risk mappings are distribution-invariant.
\begin{assumption}\label{assump:trans-inv}
    $\rho_t$ is distribution-invariant, $\forall t$.
\end{assumption}

We consider two cases: (i) the transition kernel is stochastically monotone on the entire $\mathcal X$, yielding value function monotonicity jointly in all state dimensions; and (ii) stochastic monotonicity holds only on a specified subset of dimensions conditional on the rest, implying value function monotonicity in those dimensions.

\subsection{Monotonicity in All State Dimensions}
For the case where the transition kernel is stochastically monotone on the entire $\mathcal X$, the classical risk-neutral argument carries over to the risk-averse setting \cite{smith2002structural}: if the one-step cost functions are monotone and the kernel is stochastically monotone, then the value function inherits monotonicity. We now state the assumptions and results. Let $\preceq_c$ denote the componentwise partial order on $\mathcal X\subseteq\R^n$.

\begin{assumption}\label{assump:mono-value}
   For each $t=0,\dots,T-1$: (1) $Q_t(\cdot| x)$ is stochastically increasing in $x\in\mathcal X$, and (2) $c_t(x)$ and $s_t(x)$ are decreasing in $x\in\mathcal X$.
\end{assumption}

\begin{theorem}\label{prop:mono-value}
    Under \Cref{assump:trans-inv,assump:mono-value}, $v_t(x)$ is decreasing in $x$, $\forall t$.
\end{theorem}
\begin{proof}
    We argue by backward induction. \textbf{Base case:} For $t=T$, $v_T(x)=s_T(x)$ is decreasing by assumption. \textbf{Inductive step:} Suppose $v_{t+1}(x)$ is decreasing for some $t<T$. Fix $x\preceq_c x'$. Since $Q_t(\cdot| x)$ is stochastically increasing, we have $\tilde X_{t+1}(x)\preceq_{sd}\tilde X_{t+1}(x')$. By the induction hypothesis, $v_{t+1}$ is decreasing, hence $v_{t+1}\big(\tilde X_{t+1}(x)\big) \succeq_{sd} v_{t+1}\big(\tilde X_{t+1}(x')\big)$. By \Cref{lemma:trans-inv}, $\rho_t\big(v_{t+1}\big(\tilde X_{t+1}(x)\big)\big)\geq \rho_t\big(v_{t+1}\big(\tilde X_{t+1}(x')\big)\big)$, so $x\mapsto\rho_t\big(v_{t+1}\big(\tilde X_{t+1}(x)\big)\big)$ is decreasing. Since $c_t$ and $s_t$ are decreasing, and the minimum of decreasing functions is decreasing, we conclude $v_t$ is decreasing.
\end{proof}

Given that $v_t$ is decreasing, the dual representation of the one-step risk mapping in \Cref{eq:dual-rep} provides additional insight. In general, the maximizing distribution $P^*\in \mathcal A_t(Q_t(\cdot| x_t))$ in \Cref{eq:dual-rep} depends on the particular function $v$. However, if the risk envelope admits a minimal element w.r.t. FOSD, there exists a single maximizer that works for every decreasing $v$. In this case, the coherent risk reduces to an expectation under a fixed worst-case distribution, and the risk-averse MDP becomes equivalent to a risk-neutral formulation with a modified transition law, as formalized below.
\begin{assumption}\label{assump:minimal-elem}
    The risk envelope $\mathcal A_t(Q_t(\cdot| x))$ has a minimal element $\tilde Q_t(\cdot| x)$ w.r.t. FOSD, $\forall t$ and $x\in\mathcal X$.
\end{assumption}

\begin{proposition}\label{thm:cvar-equiv}
    Under \Cref{assump:trans-inv,assump:mono-value,assump:minimal-elem}, the risk-averse MDP has the same value functions and optimal policies as a risk-neutral MDP with transition kernel $\tilde Q_t(\cdot \mid x)$ and identical cost functions.
\end{proposition}

\begin{proof}
    By \Cref{prop:mono-value}, $v_t$ is decreasing. Under \Cref{assump:minimal-elem}, by the definition of FOSD in \Cref{def:stochastic-mono}, the supremum in \Cref{eq:dual-rep} is attained at $\tilde Q_t(\cdot | x_t)$ for any decreasing function $v$. Hence, $\rho_t\big(v_t\big(\tilde X_{t+1}(x_t)\big)\big)= \E_{\tilde X_{t+1}\sim \tilde Q_t(\cdot | x)} \big(v_t\big(\tilde X_{t+1}\big)\big).$ Consequently, the DP recursion \Cref{eq:dp} becomes $v_t(x_t)=\min\big\{s_t(x_t),c_t(x_t)+\E_{\tilde X_{t+1}\sim \tilde Q_t(\cdot | x)} \big(v_t\big(\tilde X_{t+1}\big)\big)\big\}$ for $t< T$ and $v_T(x_T)=s_T(x_T)$, which is exactly the DP recursion of a risk-neutral MDP with transition kernel $\tilde Q_t(\cdot | x)$ and the same cost functions.
\end{proof}
A minimal element of the risk envelope $\mathcal A_t(Q_t(\cdot| x))$ need not exist in general. However, it can be identified when the envelope has additional order structure---for example, if $\mathcal A_t(Q_t(\cdot| x))$ forms a lattice under FOSD, as in \cite{kertz2000complete,yang2017monotone}---in which case, under suitable conditions (e.g., completeness or boundedness of the lattice), a minimal element exists. Among common coherent risk measures, CVaR provides a canonical example as illustrated below.  Such lattice structure can also be imposed---for instance, in robust MDP formulations where the ambiguity set of transition probabilities (the ``risk envelope'' in this setting) can be endowed with a lattice order \cite{wiesemann2013robust}.

\begin{example}
    Let $\rho_t=\cvar_{\alpha,t},~\forall t$. For simplicity, assume $Q_t(\cdot| x)$ admits a density $q_t(\cdot| x)$ and the state space $\mathcal X\subseteq\R$ is a compact interval (totally ordered), and consider a time-homogeneous MDP; we therefore drop the subscript $t$. The following observations extend to general transition kernels and any totally ordered $\mathcal X$: (i) $\forall x\in\mathcal X$, the CVaR risk envelope $\mathcal A(Q(\cdot | x))$ admits a minimal element w.r.t. FOSD, denoted by $\tilde Q(\cdot\mid x)$; and (ii) if $Q(\cdot| x)$ is stochastically increasing in $x$, then so is $\tilde Q(\cdot| x)$.
    We now verify (i). Since every $P\in\mathcal A(Q(\cdot| x))$ is absolutely continuous w.r.t.\ $Q(\cdot| x)$, we write its density as $p:\mathcal X\mapsto\R_+$. By \Cref{eq:dual-rep} and \Cref{exp:cvar-dual}, we can write $\rho\big(v\big(\tilde X(x)\big)\big) = \sup_{p\in \mathcal A(Q(\cdot | x))} \E_{\tilde X\sim p} v\big(\tilde X\big)$, where $\mathcal A(Q(\cdot | x))=\big\{p\in\mathcal P(\mathcal X) \ | \ p(y) \leq q(y|x)/\alpha,~\forall y\big\}$. If $v$ is decreasing, the supremum is attained by concentrating probability mass on the lower $\alpha$-tail of $Q(\cdot| x)$. Let $q_\alpha (x) := \sup\big\{ x' \ | \ \int_{-\infty}^{x'} q(y|x) dy\leq\alpha \big\}$ be the lower $\alpha$-quantile of $q(\cdot|x)$. It is straightforward to check that the maximizer---and hence the minimal element---is $\tilde q(y|x)=(q(y|x)/\alpha) \1\{y\leq q_\alpha (x)\}$, i.e., $\tilde Q(\cdot| x)$ is obtained by truncating $Q(\cdot| x)$ to its lower $\alpha$-tail and renormalizing. Consequently, for any decreasing $v$, $ \cvar_\alpha\big(v\big(\tilde X(x)\big)\big)=\E_{\tilde X\sim \tilde q(\cdot| x)}\big(v\big(\tilde X\big)\big)$. To verify (ii), fix $x'\leq x$. Then, $Q(\cdot| x')\preceq_{sd} Q(\cdot| x)$ implies $q_\alpha(x')\leq q_\alpha(x)$. Let $F_Q(\cdot|x),F_{\tilde Q}(\cdot|x)$ be the CDFs of $Q(\cdot| x),\tilde Q(\cdot| x)$, respectively. For any $q\leq q_\alpha(x')$, $F_{\tilde Q}(q|x')=F_Q(q|x')/\alpha\geq F_Q(q|x)/\alpha=F_{\tilde Q}(q|x')$, so $\tilde Q(\cdot| x')\preceq_{sd}\tilde Q(\cdot\mid x)$. \QEDB
\end{example}

\subsection{Monotonicity in a Subset of Dimensions}
For multidimensional (partially ordered) state spaces, verifying stochastic monotonicity on the full space can be challenging. In practice, it is often easier and still useful to establish monotonicity along individual dimensions. Such dynamics arise in many stochastic models (e.g., organ transplantation \cite{ren2024optimal} and option pricing \cite{wu2003optimal}), where some state components evolve monotonically when the others are held fixed. Motivated by these settings, we consider product-form state spaces.
\begin{assumption}\label{assump:prod-space}
    $\mathcal X=\mathcal X_1\times \mathcal X_2$, where $\mathcal X_1\subseteq \R^{n_1}$ is equipped with a partial order $\preceq_1$, and $\mathcal X_2\subseteq \R^{n_2}$, with $n_1+n_2=n,~n_1,n_2\geq 1,$.
\end{assumption}

\begin{remark}
    Notice that $(\preceq_1,=)$ defines a partial order on $\mathcal X_1\times \mathcal X_2$; see \cite{jiang2015approximate}.
\end{remark}

Throughout the paper, we say that a multivariable function is monotone in some
component if the function is monotone in that component with other components fixed.

\begin{assumption}\label{assump:mono-value-partial}
    Let $x=(x_1,x_2)$ with $x_i\in\mathcal X_i$ for $i=1,2$, and $\tilde X_{t}(x)=\big(\tilde X_{t,1}(x),\tilde X_{t,2}(x)\big)\sim Q_t(\cdot | x)$. For $t=0,\ldots,T$:
    \begin{itemize}
        \item $c_t(x_1,x_2)$ and $s_t(x_1,x_2)$ are decreasing in $x_1$. 
        \item $\tilde X_{t,2}(x)$ depends only on $x_2$.
        \item For fixed $x_2$ and conditioning on $\tilde X_{t,2}(x)$, $\tilde X_{t,1}(x_1,x_2)$ is stochastically increasing in $x_1$, i.e., $x_1\preceq_1 x_1'\implies\tilde X_{t,1}(x_1,x_2) \preceq_{sd} \tilde X_{t,1}(x_1',x_2)$.
    \end{itemize}
\end{assumption}
By the second bullet of \Cref{assump:mono-value-partial}, we may drop the dependence on $x_1$ and write $\tilde X_{t,2}(x_2)$. Similar (sometimes slightly weaker) coordinate-wise conditions have been used to derive monotonicity in $\mathcal X_1$ of performance functions in risk-neutral settings; see, e.g., \cite{oh2016characterizing,ren2024optimal,jiang2015approximate}. The usual risk-neutral argument applies the tower property by conditioning on $\tilde X_{t,2}(x_2)$, e.g., expressing
\begin{align*}
    \E\big(v\big(\tilde X_{t}(x)\big)\big)=\E\big(\E\big(v\big(\tilde X_{t,1}(x),\tilde X_{t,2}(x_2)\big)|\tilde X_{t,2}(x_2)\big)\big),
\end{align*}
then shows the inner conditional expectation is monotone in $x_1$. This approach does not extend to risk-averse settings because the tower property fails. Instead, the proof of the next theorem, which establishes monotonicity in $\mathcal X_1$, uses a coupling and stochastic dominance argument that circumvents this issue \cite{strassen1965existence}.
\begin{theorem}\label{thm:mono-value-partial}
    Under \Cref{assump:trans-inv,assump:prod-space,assump:mono-value-partial}, $v_t(x_1,x_2)$ is decreasing in $x_1$, $\forall t$.
\end{theorem}

\begin{proof}
    We argue by backward induction on $t$. \textbf{Base case:} For $t=T$, $v_T(x_1,x_2)=s_T(x_1,x_2)$ is decreasing in $x_1$ by assumption. \textbf{Inductive step}: Fix $t<T$ and assume $v_{t+1}(x_1,x_2)$ is decreasing in $x_1$ for each fixed $x_2$. Let $x=(x_1,x_2)$ and $x'=(x_1',x_2)$ with $x_1\preceq_1 x_1'$. By \Cref{assump:mono-value-partial}, we have $\tilde X_{t+1,2}(x)=\tilde X_{t+1,2}(x')$ a.s., and $\tilde X_{t+1,1}(x) \preceq_{sd} \tilde X_{t+1,1}(x')$. Construct next-state vectors $\hat X_{t+1}(x)=\big(\hat X_{t+1,1}(x),\hat X_{t+1,2}(x_2)\big)$, $\hat X_{t+1}(x')=\big(\hat X_{t+1,1}(x'),\hat X_{t+1,2}(x_2)\big)$ as follows: (i) Synchronize the second coordinate by taking $\hat X_{t+1,2}(x_2)=\tilde X_{t+1,2}(x_2)$ a.s.; (ii) Conditioning on $\hat X_{t+1,2}(x_2)$, consider the first coordinate. Under \Cref{assump:mono-value-partial} (third bullet), by Strassen’s theorem \cite{strassen1965existence,lindvall1999strassen}, there exists random variables $\hat X_{t+1,1}(x)$ and $\hat X_{t+1,1}(x')$ such that (i) $\hat X_{t+1,1}(x)\stackrel{d}{=}\tilde X_{t+1,1}(x)$, (ii) $\hat X_{t+1,1}(x')\stackrel{d}{=}\tilde X_{t+1,1}(x')$, and (iii) $\hat X_{t+1,1}(x)\preceq_1 \hat X_{t+1,1}(x')$ a.s. Consequently, $\hat X_{t+1}(x)\stackrel{d}{=}\tilde X_{t+1}(x)$, $\hat X_{t+1}(x')\stackrel{d}{=}\tilde X_{t+1}(x')$, and $\hat X_{t+1,2}(x) = \hat X_{t+1,2}(x')$, $\hat X_{t+1,1}(x) \preceq_{sd} \hat X_{t+1,1}(x')$ a.s. 
    Since $v_{t+1}(x_1,x_2)$ is decreasing in $x_1$, we have $v_{t+1}\big(\hat X_{t+1}(x)\big) \geq v_{t+1}\big(\hat X_{t+1}(x')\big)$ a.s., and thus $\rho_t(v_{t}(\hat X_{t}(x)))\geq \rho_t(v_{t}(\hat X_{t+1}(x')))$. Since $\rho_t$ is distribution-invariant, by \Cref{lemma:trans-inv}, $\rho_t\big(v_{t+1}(\tilde X_{t+1}(x))\big)\geq \rho_t\big(v_{t+1}(\tilde X_{t+1}(x'))\big)$, so $\rho_t\big(v_{t+1}(\tilde X_{t+1}(x))\big)$ is decreasing in $x_1$. Finally, since $s_t(\cdot,x_2)$ and $c_t(\cdot,x_2)$ are decreasing for fixed $x_2$ and the minimum of decreasing functions is decreasing, the DP recursion \eqref{eq:dp} implies that $v_{t}(x_1,x_2)$ is decreasing in $x_1$.
\end{proof}

The following average-rate forward (ARF) model with an early-termination feature illustrates \Cref{thm:mono-value-partial}. The Asian–American call option has the same two-dimensional state dynamics as \Cref{exp:a-a-option} but a slightly different option-style terminal payoff \cite{wu2003optimal}; the analysis and results below carry over. 
\begin{example}\label{exp:a-a-option}
    An ARF is a foreign-exchange derivative whose settlement depends on the arithmetic average of the spot over a specified window. Consider a discrete-time ARF with the possibility of early termination and state space $\mathcal X=\R^2_{+}$. Let $X_t=(X_{t,1},X_{t,2})$, where $X_{t,2}$ is the current spot and $X_{t,1}$ is the running average rate up to time $t$. Let $\{W_t\}$ be an independent and identically distributed (i.i.d.) sequence of log-normal random variables (a single scalar shock). The next state $\tilde X_{t+1}(x_t)$ is given by
    \begin{align*}
        \begin{cases}
            \tilde X_{t+1,1}(x_t)=\big(t x_{t,1}+\tilde X_{t+1,2}(x_t)\big)/(t+1),\\
            \tilde X_{t+1,2}(x_t)=W_t x_{t,2}.
        \end{cases}
    \end{align*}
    At time $t$, terminating the ARF delivers an early-termination settlement $a_t x_{t,1}+ b_t x_{t,2} + c_t$, with $a_t,b_t>0$ and $c_t\in\R$ the contract's delivery (strike) rate. The stopping cost is the negative of this early-termination settlement, $s_t(x_t)=-(a_t x_{t,1}+ b_t x_{t,2} + c_t)$; there is no running cost. Observe that, holding $X_{t+1,2}(x_t)$ fixed, $\tilde X_{t+1,1}(x)$ is increasing and therefore stochastically increasing in $x_1$. On the other hand, $\tilde X_{t+1,2}(x)$ depends only on $x_2$. By \Cref{thm:mono-value-partial}, $v_t(x)$ is decreasing in $x_1$. 
    
    Indeed, in this example, $v_t(x)$ is decreasing in each of the coordinates; this also follows from \Cref{prop:mono-value}, since $\tilde X_{t+1}(x_t)$ is stochastically increasing in $x_t$ under the componentwise partial order on $\R^2$. \QEDB
\end{example}

For another example satisfying \Cref{assump:mono-value-partial}, see the organ transplantation model in \cite{ren2024optimal}, where $X_t=(X_{t,1},X_{t,2})$: $X_{t,2}$ is the patient’s health state, evolving only from the previous health state, and $X_{t,1}$ is the organ-offer quality state, which is stochastically monotone in the health state (patients in better health are more likely to receive higher-quality offers).


\section{Characterization of the structure of the optimal policy}\label{sec:control-limit}
In this section, we establish conditions for the existence of control limit optimal policies, beginning with a precise definition. Let $x=(x_1,\cdots,x_n)\in\mathcal X\subseteq\R^n$ and define the $(n-1)$-dimensional vector $x_{-i}$ to be the vector $x$ with its $i$-th component removed. Throughout this section, we consider monotonicity w.r.t. the componentwise partial order $\preceq_c$ on $\mathcal X\subseteq\R^n$.
\begin{definition}
    A policy $d=(d_0,\cdots,d_{T-1})\in\mathcal D$ is a \textit{control limit policy} in the $i$-th dimension if, for each $t=0,\ldots,T-1$, there exists a threshold function $\overline{x}_t^i:\R^{n-1}\mapsto\R\cup\{\pm\infty\}$ such that for any $x\in\mathcal X$,
    \begin{align*}
        d_t(x)=\begin{cases}
            C   &x_{i}\geq \overline{x}_t^i(x_{-i}) ~(\text{resp., } \leq),\\
            S   &x_{i}< \overline{x}_t^i(x_{-i}) ~(\text{resp., } >).
        \end{cases}
    \end{align*}
    The inequality orientation and the weak/strict boundary convention may be reversed. When $\mathcal X$ is totally ordered, the dependence of $\overline{x}_t^i(x_{-i})$ on $x_{-i}$ can be dropped.
\end{definition}

We provide a general framework to characterize policy structure. For each $t$ and $x_t\in\mathcal X$, define the \textit{continuation loss}
\begin{align}\label{eq:cont-loss}
    L_t(x_t):=\rho_t\big(v_{t+1}\big(\tilde X_{t+1}(x_t)\big)\big)+c_t(x_t)-s_t(x_t).
\end{align}
$L_t(x_t)$ measures the increased risk from delaying termination at time $t$. By the DP recursion \eqref{eq:dp}, the optimal action is determined by the sign of $L_t(x_t)$; we therefore have the following result (proof omitted).
\begin{lemma}\label{lemma:cl-policy}
    \begin{itemize}
        \item Continuation $C$ is optimal at time $t$ if and only if (iff) $L_t(x_t)\leq 0$.
        \item For each $t$ and some fixed $i$, if $L_t(x_t)$ is decreasing (resp., increasing) in $x_{t,i}$ for each fixed $x_{t,-i}$, a control limit optimal policy in the $i$-th dimension exists: there exists a threshold function $\overline{x}_t^i:\R^{n-1}\mapsto\R\cup\{\pm\infty\}$ such that it is optimal to continue (resp., stop) at time $t$ iff $x_{t,i}\geq \overline{x}_t^i(x_{t,-i})$.
    \end{itemize}
\end{lemma}

The result above aligns with classical conditions for general MDPs to admit monotone optimal policies (which reduce to control limit policies when $|\mathcal A|=2$). To see this, define the risk-to-go
\begin{align*}
    Q_t(x,a)=\begin{cases}
        s_t(x), &a=S,\\
        \rho_t\big(v_{t+1}\big(\tilde X_{t+1}(x_t)\big)\big)+c_t(x_t), &a=C.\\
    \end{cases}
\end{align*}
It was established in \cite{topkis1978minimizing,serfozo1976monotone} that a monotone optimal policy exists if $Q_t(x,a)$ has antitone or isotone differences, i.e., $Q_t(\cdot,C)-Q_t(\cdot,S)$ is monotone, which in our optimal stopping setting is exactly the condition that $L_t(x)$ be monotone. To enforce this monotonicity, work on monotone policies in general MDPs \cite{puterman2014markov,lovejoy1987some,miehling2020monotonicity,brau1997controlled} typically imposes separate modularity conditions on the transition kernel and on costs. In our setting (consider the case $L_t$ is decreasing), these respectively imply the following conditions:
\begin{enumerate}
    \item[(C1)] $\rho_t\big(v_{t+1}\big(\tilde X_{t+1}(x_t)\big)\big)$ is decreasing in $x_t$,
    \item[(C2)] $c_t(x_t)-s_t(x_t)$ is decreasing in $x_t$.
\end{enumerate}
While (C1) is fairly common---indeed, it is established in the proof of \Cref{prop:mono-value}---(C2) is often violated in optimal stopping. In \Cref{assump:mono-value}, both $c_t$ and $s_t$ are decreasing, but $c_t$ is an intermediate (one-period) cost, whereas $s_t$ is a terminal cost that aggregates long-term effects and typically dominates $c_t$. Consequently, $c_t(x_t)-s_t(x_t)$ is dominated by $-s_t(x_t)$, which is increasing. For example, in the organ transplantation model of \cite{ren2024optimal}, $c_t(x_t)$ may equal the length of a decision period (in weeks), while $s_t(x_t)$ is post-transplant life expectancy (in years), often an order of magnitude (tenfold or more) larger.

For risk-neutral optimal stopping, \cite{oh2016characterizing} proposed a practical criterion to certify the existence of control limit optimal policies; however, it fails in risk-averse settings because expectation is additive whereas coherent risk measures are only subadditive. To highlight this gap, we adapt their results to the risk-averse setting. Define the \textit{one-step loss}
\begin{align*}
    M_t(x_t):=\rho_t\big(s_{t+1}\big(\tilde X_{t+1}(x_t)\big)\big)+c_t(x_t)-s_t(x_t),
\end{align*}
which captures the loss incurred when postponing termination from $t$ to $t+1$. The relationship between $L_t$ and $M_t$ is given by the following result.
\begin{proposition}\label{prop:cl-frame}
    \begin{itemize}
        \item For each $t$ and $x_t\in\mathcal X$, $L_t(x_t)\leq M_t(x_t)$.
        \item If $Q_t(\cdot| x)$ is stochastically increasing in $x\in\mathcal X$ and $\rho_t$ is additive for each $t$, then $M_t$ decreasing (in some fixed coordinate) implies that $L_t$ is decreasing (in the same coordinate).
    \end{itemize}
\end{proposition}
\begin{proof}
    To see that $L_t(x_t)\leq M_t(x_t)$, observe that
    \begin{align}\label{eq:subadd}
        \begin{split}
        L_t(x_t)&=\rho_t\big(v_{t+1}\big(\tilde X_{t+1}(x_t)\big)\big)+c_t(x_t)-s_t(x_t)\\
        &=\rho_t\big(\min\big\{s_{t+1}\big(\tilde X_{t+1}(x_t)\big),s_{t+1}\big(\tilde x_{t+1}(x_t)\big)\\
        &+L_{t+1}\big(\tilde X_{t+1}(x_t)\big) \big\}\big)+c_t(x_t)-s_t(x_t)\\
        &\leq M_t(x_t) + \rho_t\big(\min\big\{0,L_{t+1}\big(\tilde X_{t+1}(x_t)\big) \big\}\big)\\
        &\leq M_t(x_t).
        \end{split}
    \end{align}
    where the first and second inequalities follow from subadditivity and monotonicity of $\rho_t$, respectively. If $\rho_t$ is additive for each $t$, the first inequality becomes an equality, and the risk-neutral backward induction argument in \cite[Section 2]{oh2016characterizing} applies and proves that $M_t$ decreasing implies that $L_t$ is decreasing.
\end{proof}

\Cref{prop:cl-frame} indicates that in the risk-neutral case (each $\rho_t$ is a conditional expectation), if $Q_t(\cdot| x)$ is stochastically increasing in $x\in\mathcal X$, the condition that $M_t$ is decreasing suffices for the existence of control limit optimal policies. Moreover, this condition is preferable to the classical modularity conditions (C1) and (C2). First, it is easier to check, since $M_t$ depends only on primitives (cost functions, the transition law, and the one-step risk mappings), whereas establishing (C1) is more challenging. Second, it is weaker, as illustrated below.
\begin{example}
    Consider a time-homogeneous risk-neutral MDP where each $\rho_t$ is a conditional expectation, so we can drop time subscripts on the cost functions, risk functionals, and transition law. Then (C1) implies that $\rho\big(s\big(\tilde X(x)\big)\big)=\rho\big(v_{T}\big(\tilde X(x)\big)\big)$ is decreasing in $x$, which, together with (C2), \textit{implies} that $M_t$ is decreasing. \QEDB
\end{example}


However, since $\rho_t$ is subadditive, in the general risk-averse case, $L_t$ no longer inherits the monotonicity of $M_t$. Guided by \Cref{prop:cl-frame} and the analysis that follows, we therefore seek conditions in risk-averse settings under which the structure of $M_t$ still implies an optimal policy structure. In \Cref{sec:policy-mono1}, we establish that if both the one-step risk mappings and the state vector satisfy appropriate \textit{comonotonicity} conditions, subadditivity tightens to additivity, restoring the risk-neutral argument. In \Cref{sec:policy-mono2}, we give conditions under which the sign of $L_t$ is determined entirely by $M_t$; consequently, a \textit{one-step look-ahead} policy is optimal.

\subsection{Comonotone Risk Measures and State Vectors}\label{sec:policy-mono1}
In this section, we focus on \textit{comonotone} risk measures, a special class of risk measures that are additive for comonotone random variables. We begin by introducing the notion of comonotonicity.
\begin{definition}
    \begin{itemize}
        \item Random variables $X,Y$ on measurable space $(\Omega,\mathcal F)$ are comonotone if $(X(\omega)-X(\omega'))(Y(\omega)-Y(\omega'))\geq 0,~\forall\omega,\omega'\in\Omega$.
        \item An $n$-dimensional random vector $X=(X_1,\cdots,X_n)$ on $(\Omega,\mathcal F)$ is comonotone if its components are pairwise comonotone.
    \end{itemize}
\end{definition}
Comonotone random variables have the following equivalent characterizations \cite{puccetti2010multivariate}.
\begin{proposition}\label{prop:como=mono-fun}
    Let $X,Y$ be random variables. Then, $X,Y$ are comonotone iff there exists a random variable $Z$ and increasing functions $f,g:\R\mapsto\R$ such that $(X,Y)\stackrel{d}{=}(f(Z),g(Z))$. In particular, without loss of generality, one may take $Z\sim\text{Uniform}(0,1)$, and $f=F^{-1}_{X},~g=F^{-1}_{Y}$, where $F_{X},F_{Y}$ are the marginal CDFs of $X,Y$, respectively.    
\end{proposition}
\begin{definition}
    A risk measure $\rho$ is comonotone if $\rho(X+Y)=\rho(X)+\rho(Y)$ for any random variables $X,Y$ that are comonotone. 
\end{definition}
Reference \cite{kusuoka2001law} provides an explicit form for comonotone coherent risk measures: a coherent risk measure is distribution-invariant, comonotone, and has the Fatou property iff it admits a representation as an integral of the quantile function w.r.t. a positive measure. Spectral risk measures---of which CVaR is a special case---are canonical examples of this integral form. 

Inspecting \Cref{eq:subadd}, if $\rho_t$ is comonotone and $s_{t+1}\big(\tilde X_{t+1}(x_t)\big)$ and $\min\big\{0, L_{t+1}\big(\tilde X_{t+1}(x_t)\big)\big\}$ are comonotone, then the first inequality holds with equality, restoring the risk-neutral inductive step that $L_t$ inherits the monotonicity of $M_t$. We therefore assume:
\begin{assumption}\label{assump:como-measure}
    For each $t=0,1,\cdots,T-1$, $\rho_t$ is a comonotone risk measure.
\end{assumption}
It remains to ensure the comonotonicity of $s_{t+1}\big(\tilde X_{t+1}(x_t)\big)$ and $\min\big\{0, L_{t+1}\big(\tilde X_{t+1}(x_t)\big)\big\}$. Since the monotonicity of $L_{t+1}$ can be established inductively and that of $s_{t+1}$ is typically assumed, we seek conditions guaranteeing this comonotonicity whenever $L_{t+1}$ and $s_{t+1}$ are monotone (in the same direction). We impose the following condition on the system dynamics:
\begin{assumption}\label{assump:como-vector}
    The random vector $\tilde X_{t+1}(x_t)$ is comonotone, $\forall t$ and $x_t\in\mathcal X$.
\end{assumption}
\Cref{assump:como-vector} implies that the components of the state vector are positively dependent. Notice that when $\mathcal X$ is a totally ordered state space (e.g., $\mathcal X\subseteq\R$), \Cref{assump:como-vector} holds trivially. The next example describes a class of multidimensional system dynamics for which \Cref{assump:como-vector} holds, with the ARF model in \Cref{exp:a-a-option} as a special case.
\begin{example}\label{exp:illus-como-vector}
    Consider an optimal stopping problem with state vectors $\{X_t\}$ driven by an independent sequence of random variables $\{W_t\}$ as follows: $X_{t+1}(x_t)=f_t(x_t,W_{t})$, where $f_t:\mathcal X\times \R\mapsto \mathcal X$. Writing $f_t=(f_{t,1},\cdots,f_{t,n})$, assume that for any $x\in\mathcal X$ and each $i$, the map $f_{t,i}(x,\cdot):\R\mapsto\R$ is monotone in the same direction. Thus every component of the next state $\tilde X_{t+1}(x_t)$ is a monotone function of the same exogenous shock $W_t$, so by \Cref{prop:como=mono-fun}, $\tilde X_{t+1}(x_t)$ is comonotone. 
    
    The ARF model in \Cref{exp:a-a-option} fits this setup and can be written as
    \begin{align*}
        \begin{cases}
            \tilde X_{t+1,1}(x_t)=(t x_{t,1}+W_t x_{t,2})/(t+1),\\
            \tilde X_{t+1,2}(x_t)=W_t x_{t,2},
        \end{cases}
    \end{align*}
    where both components are increasing in $W_t$. \QEDB
\end{example}

\Cref{assump:como-measure,assump:como-vector}, together with the following monotonicity assumption on $M_t$ and $s_t$, allow us to establish the existence of a control limit optimal policy in every dimension.
\begin{assumption}\label{assump:threshold-1d}
    For each $t=0,\cdots,T-1$, $M_t(x)$ and $s_t(x)$ are decreasing in $x$.
\end{assumption}
\begin{theorem}\label{thm:threshold-1d}
    Under \Cref{assump:trans-inv,assump:como-measure,assump:como-vector,assump:threshold-1d}, for each $t=0,\cdots,T-1$, $L_t(x)$ is decreasing in $x$, and thus a control limit optimal policy in every dimension exists.
\end{theorem}
\begin{proof}
    We argue by backward induction on $t$. \textbf{Base case}: By definition, $L_{T-1}(x)=M_{T-1}(x)$, which is decreasing by \Cref{assump:threshold-1d}. \textbf{Induction step}: Suppose $L_{t+1}$ is decreasing. Then,
    \begin{align}\label{eq:como-proof}
        \begin{split}
            &~~~~\rho_t\big(v_{t+1}\big(\tilde X_{t+1}(x_t)\big)\big)\\
            &=\rho_t\big(\min\big\{s_{t+1}\big(\tilde X_{t+1}(x_t)\big),s_{t+1}\big(\tilde x_{t+1}(x_t)\big)\\
            &+L_{t+1}\big(\tilde X_{t+1}(x_t)\big) \big\}\big)\\
            &=\rho_t\big(s_{t+1}\big(\tilde X_{t+1}(x_t)\big)+\min\big\{0,L_{t+1}\big(\tilde X_{t+1}(x_t)\big)\big\}\big).
        \end{split}
    \end{align}
    Since $L_{t+1}$ is decreasing, $\min\{0,L_{t+1}(\cdot)\}$ is also decreasing. Fix $x_t\in\mathcal X$. By \Cref{assump:como-vector} and \Cref{prop:como=mono-fun}, there exists a random variable $Y_t$ and a mapping $f: \R\mapsto \mathcal X$ with each component $f_i:\R\mapsto\R$ increasing such that $\tilde X_{t+1}(x_t)=f(Y_t)$. Hence the compositions $s_{t+1}(f(\cdot)),\min\{0,L_{t+1}(f(\cdot))\}$ are decreasing. By \Cref{prop:como=mono-fun}, $s_{t+1}\big(\tilde X_{t+1}(x_t)\big)$ and $\min\big\{0,L_{t+1}\big(\tilde X_{t+1}(x_t)\big)\big\}$ are comonotone. Since $\rho_t$ is comonotone, \eqref{eq:como-proof} yields
    \begin{align*}
        &\rho_t\big(v_{t+1}\big(\tilde X_{t+1}(x_t)\big)\big)\\
        &=\rho_t\big(s_{t+1}\big(\tilde X_{t+1}(x_t)\big)\big)+\rho_t\big(\min\big\{0,L_{t+1}\big(\tilde X_{t+1}(x_t)\big) \big\}\big),
    \end{align*}
    so the inequality in \Cref{eq:subadd} holds with equality:
    \begin{align*}
        L_t(x_t)= M_t(x_t) + \rho_t(\min\{0,L_{t+1}(\tilde X_{t+1}(x_t)) \}).
    \end{align*}
    By \Cref{assump:threshold-1d}, $M_t$ is decreasing; combined with the monotonicity of $\rho_t$ and the inductive hypothesis, this implies $L_t$ is decreasing.
\end{proof}

The following two examples, with partially ordered and totally ordered state spaces respectively, illustrate \Cref{thm:threshold-1d}.
\begin{example}
    Consider the ARF model from \Cref{exp:a-a-option}. As shown in \Cref{exp:illus-como-vector}, \Cref{assump:como-vector} holds. Suppose \Cref{assump:trans-inv,assump:como-measure} also hold. To establish the existence of control limit optimal policies, it remains to verify \Cref{assump:threshold-1d}. The stopping cost $s_t(x_t)=-(a_t x_{t,1}+b_t x_{t,2} + c_t)$ is decreasing in both coordinates. For $M_t$, we compute 
    \begin{align*}
        &~~~~M_t(x_t)\\
        &=\rho_t\left(-\left(a_{t+1} \cdot \frac{t x_{t,1}+W_t x_{t,2}}{t+1}+b_{t+1} W_t x_{t,2}+c_{t+1}\right)\right)\\
        &+(a_t x_{t,1}+b_t x_{t,2} + c_t)\\
        &= \left(\rho_t \left( -\left(\frac{a_{t+1}}{t+1}+b_{t+1} \right)W_t\right)+b_t\right)x_{t,2}\\
        &+\left(-\frac{ta_{t+1}}{t+1}+a_t\right)x_{t,1}+c_t-c_{t+1}.
    \end{align*}
    Hence, if the coefficients of $x_{t,1}$ and $x_{t,2}$ above are nonpositive, then $M_t$ is decreasing in both coordinates, and control limit optimal policies exist in both coordinates. \QEDB
\end{example}
\begin{example}[Selling with a deadline \cite{bertsekas2012dynamic}]
    Consider selling a required quantity of raw material before a fixed deadline. The material price fluctuates over time, and the decision is whether to sell immediately at the current price or wait for a later period. Suppose the price process $\{X_t\}$ follows $X_{t+1}=\lambda X_t + W_t$, where $\{W_t\}$ is i.i.d. and $ \lambda > 1$ is constant. The stopping cost is the negative current price, $s_t(x)=-x$, with no running cost. The one-step loss is
    \begin{align*}
        M_t(x_t)=\rho_t(-X_{t+1}(x_t))+x_t=(-\lambda+1)x_t + \rho_t(-W_t),
    \end{align*}
    which is strictly decreasing in $x_t$. Hence, if $\rho_t$ is distribution-invariant and comonotone, a control limit optimal policy exists. \QEDB
\end{example}

\subsection{One-Step Look-Ahead Optimal Policy}\label{sec:policy-mono2}
We define a one-step look-ahead policy $\overline d=\{\overline d_0,\cdots,\overline d_{T-1}\}$ as follows. For each $t$, the action is determined by the sign of the one-step loss $M_t$:
\begin{align*}
        \overline d_t(x_t)=\begin{cases}
        S & M_t(x_t)\geq 0,\\
        C & M_t(x_t)< 0,
    \end{cases}
\end{align*}
i.e., stopping at time $t$ whenever deferring termination to $t+1$ increases risk. Because this policy is far simpler to compute than solving the DP, it is desirable to know when it is optimal. \Cref{prop:cl-frame} states that the optimal action is characterized by the sign of $L_t$, so the one-step look-ahead policy $\overline d$ is optimal iff $M_t$ and $L_t$ have the same sign. We now give a sufficient condition.
\begin{assumption}\label{assump:one-step}
    for each $t$ and $x_t\in\mathcal X$, if $M_t(x_t)\geq 0$, then $M_{t+1}\big(\tilde X_{t+1}(x_t)\big)\geq 0$ a.s.
\end{assumption}
\begin{theorem}\label{thm:one-step}
    Under \Cref{assump:trans-inv,assump:one-step}, $M_t$ and $L_t$ have the same sign, $\forall t$. Consequently, the one-step look-ahead policy $\overline d$ is optimal.
\end{theorem}
\begin{proof}
    We argue by backward induction on $t$. \textbf{Base case}: $M_{T-1}(x_{T-1})=L_{T-1}(x_{T-1})$, so they have the same sign. \textbf{Induction step}: Assume $\{x\in\mathcal X | M_{t+1}(x)\leq 0\}=\{x\in\mathcal X | L_{t+1}(x)\leq 0\}$ for some $t$. Fix $x_t\in\mathcal X$ with $M_t(x_t)\geq 0$. By \Cref{assump:one-step}, $M_{t+1}\big(\tilde X_{t+1}(x_t)\big)\geq 0$ a.s.; hence, by the induction hypothesis, $L_{t+1}\big(\tilde X_{t+1}(x_t)\big)\geq 0$ a.s. Using \Cref{eq:subadd},
    \begin{align*}
        L_t(x_t)&=\rho_t(s_{t+1}(\tilde X_{t+1}(x_t))+\min\{0,L_{t+1}(\tilde X_{t+1}(x_t))\})\\
        &+c_t(x_t)-s_t(x_t)\\
        &=\rho_t(s_{t+1}(\tilde X_{t+1}(x_t)))+c_t(x_t)-s_t(x_t)\\
        &=M_t(x_t)\geq 0.
    \end{align*}
    Conversely, by the second inequality in \Cref{eq:subadd}, $L_t(x_t)\leq M_t(x_t),~\forall x_t$. Therefore, $M_t(x_t)\leq 0\implies L_t(x_t)\leq 0$. Thus, $M_t$ and $L_t$ have the same sign, and the one-step look-ahead policy $\overline d$ is optimal at time $t$.
\end{proof}

By \Cref{thm:one-step}, if, in addition, $M_t$ is monotone for each $t$, then the one-step look-ahead policy $\overline d$ is optimal and of control limit-type. We next present verifiable sufficient conditions that ensure \Cref{assump:one-step} given $M_t$ is monotone.
\begin{assumption}\label{assump:one-step-threshold} For each $t$: (i) $M_t(x)$ is increasing (resp., decreasing) in $x$; (ii) $\forall x$, $M_t(x)$ and $M_{t+1}(x)$ have the same sign; and (iii) $x\preceq_c\tilde X_{t+1}(x)$ (resp., $\succeq_c$) a.s.
\end{assumption}
Condition (ii) is automatic in time-invariant systems. Condition (iii) captures ``non-improving'' dynamics. For instance, in transplant decision-making or machine maintenance models where larger state $X_t$ denotes poorer health, $x\preceq_c\tilde X_{t+1}(x)$ a.s. indicates that, upon continuation, the state almost surely does not improve.
\begin{corollary}\label{coro:one-step-threshold}
    Under \Cref{assump:trans-inv,assump:one-step-threshold}, the one-step look-ahead policy $\overline d$ is optimal and is a control limit policy.
\end{corollary}
\begin{proof}
    Consider the case $M_t$ is increasing (the decreasing case is analogous with reversed inequalities). Fix $x$ with $M_t(x)\geq0$. By (i) and (iii) in \Cref{assump:one-step-threshold}, $M_t\big(\tilde X_{t+1}(x)\big)\geq 0$ a.s. By (ii), $M_{t+1}\big(\tilde X_{t+1}(x)\big)\geq 0$ a.s., so \Cref{thm:one-step} implies $\overline d$ is optimal. Finally, the monotonicity of $M_t(x)$ implies a control limit structure (as in \Cref{lemma:cl-policy}).
\end{proof}

The following example illustrates \Cref{coro:one-step-threshold}. 
\begin{example}[Asset selling with past offers retained \cite{bertsekas2012dynamic}]
    An asset receives monetary offers each period. Let the offers be an i.i.d. sequence of random variables $\{W_t\}$ supported on a bounded, nonnegative interval. If an offer is accepted, the proceeds can be invested at a fixed interest rate $r$, and past offers remain available for acceptance later. Let $X_t$ be the maximum offer observed up to time $t$; then $X_{t+1}=\max\{X_t,W_t\}$. The stopping cost equals the negative of the accepted amount compounded over the remaining horizon, $s_t(x_t)=-x_t(1+r)^{T-t}$, with no running cost. The one-step loss is
    \begin{align*}
        M_t(x_t)
        &=(1+r)^{T-t-1}\rho_t(-\max\{x_t,W_t\})+(1+r)^{T-t}x_t\\
        &=(1+r)^{T-t-1}\rho_t(\min\{rx_t,(1+r)x_t-W_t\}).
    \end{align*}
    It is straightforward to verify that \Cref{assump:one-step-threshold} holds. Hence, by \Cref{coro:one-step-threshold}, $\overline d$ is optimal and of control limit-type. \QEDB
\end{example}

\begin{remark}
    In general, since $L_t(x_t)\leq M_t(x_t),~\forall x_t\in\mathcal X$, $M_t(x_t)\leq 0$ guarantees that continuation is optimal, whereas $M_t(x_t)\geq 0$ is inconclusive. If, for each $t$, $M_t$ is increasing (resp., decreasing) in its $i$-th coordinate, then for any fixed $x_{t,-i}$, there exists a threshold $\overline{x}_t^i(x_{t,-i})\in\R$ such that whenever $x_{t,i}\leq \overline{x}_t^i(x_{t,-i})$ (resp., $\geq$), $M_t(x_t)\leq 0$, and thus continuation is optimal. On the opposite side of the threshold, $M_t$ does not determine the optimal action. This resembles the $(s,S,A,p)$ policy structure in \cite{chen2007risk}.
\end{remark}
\section{Monotonicity of the optimal control limits}
In this section, we present additional monotonicity results for the optimal control limits, assuming control limit optimal policies exist. For a control limit optimal policy in the $i$-th dimension, denote the optimal control limit function by $\overline x_{t,i}:\R^{n-1}\mapsto\R$, with $\overline x_{t,i}(x_{t,-i})$ the control limit at time $t$ given $x_{t,-i}$. The statements below either parallel their risk-neutral counterparts or are immediate; proofs are omitted. We focus on the case where $L_t(x)$ is decreasing in $x$; the increasing case follows by flipping the monotonicity.
\begin{itemize}
    \item \textbf{Cross-monotonicity of state-dependent control limits.} Fix indices $i\neq j$. If $L_t(x_t)$ is decreasing in both $x_{t,i}$ and $x_{t,j}$, then control limit optimal policies exist in both dimensions, and the control limits satisfy: $\overline x_{t,i}(x_{t,-i})$ is decreasing in $x_{t,j}$ and $\overline x_{t,j}(x_{t,-j})$ is decreasing in $x_{t,i}$.
    \item \textbf{Monotonicity in time.} Consider a time-homogeneous model (the transition kernel and cost functions are fixed, and the one-step risk mappings have the same functional form across time). If $L_t(x)$ is decreasing in $x_i$ for each $t$ and some fixed $i$, then a control limit optimal policy in the $i$-th dimension exists, and the optimal control limit $\overline x_{t,i}(x_{t,-i})$ is increasing in $t$ for any fixed $x_{t,-i}$.
    \item \textbf{Monotonicity in risk-aversion level.} Consider two time-homogeneous risk-averse optimal stopping instances with identical transition kernels and cost functions, and one-step risk mappings $\{\rho_t^1\}$ and $\{\rho_t^2\}$, respectively. Let the corresponding value functions be $v_t^1$ and $v_t^2$. For some fixed $i$, suppose a control limit optimal policy in the $i$-th dimension exists in both instances, with control limit functions denoted by $\overline x_{t,i}^1$ and $\overline x_{t,i}^2$, respectively. If the first instance is more risk-averse, i.e., $\rho_t^1(Z)\geq \rho_t^2(Z)$ for each $t$ and $Z\in\mathcal Z_{t+1}$, then $v_t^1\geq v_t^2$ and $\overline x_{t,i}^1(x_{t,-i})\geq \overline x_{t,i}^2(x_{t,-i})$ for each $t$ and $x_{t,-i}$.
\end{itemize}
The first two statements mirror their risk-neutral counterparts and follow by analogous arguments; see \cite{oh2016characterizing} and \cite[Section 3.4]{bertsekas2012dynamic}. Both are intuitive. For the first statement, if the loss function is decreasing in $x_{t,i}$, the relative risk of continuing (versus stopping) falls as $x_{t,i}$ increases, so the continuation region expands, i.e., the optimal control limit $\overline x_{t,j}(x_{t,-j})$ decreases in $x_{t,i}$. For the second statement, the risk-neutral argument in \cite[Section 3.4]{bertsekas2012dynamic} carries over to the risk-averse setting and shows that $v_t(x)$---and hence $L_t(x)$---is increasing in $t$ for each fixed $x$, which yields the claim. 

The third statement formalizes the effect of increased risk aversion under identical dynamics and costs. Its proof is a straightforward backward-induction argument and is omitted. For illustration, consider one-step mean–CVaR mappings $\rho_t^i(\cdot)=(1-\alpha_i) \E(\cdot|\mathcal F_t)+\alpha_i\cvar_{\gamma,t}(\cdot)$ (with fixed $\gamma\in[0,1]$) or pure CVaR mappings $\rho_t^i(\cdot)=\cvar_{\alpha_i,t}(\cdot)$ for $i=1,2$. If $0\leq \alpha_2\leq \alpha_1 \leq1$, then $\rho_t^1(Z)\geq \rho_t^2(Z),~\forall Z\in\mathcal Z_{t+1}$. Intuitively, $\rho_t^1$ places more weight on the tail or on a more extreme tail and is therefore more risk-averse. Consequently, $v_t^1\geq v_t^2$, and the more risk-averse instance tends to stop earlier.

\section{Conclusions}\label{sec:conc}
In this paper, we establish structural results for finite-horizon optimal stopping under time-consistent dynamic coherent risk measures. Because coherent risk measures are subadditive and generally do not satisfy the tower property, risk-neutral results do not carry over directly. We show that the value function is monotone under conditions paralleling the risk-neutral case, with proofs adapted to use coupling arguments in place of conditioning-based techniques that rely on the tower property. We also develop a general framework for establishing control limit optimal policies in risk-averse settings and clarify how it differs from the standard framework for proving monotone policies in general MDPs. Within this framework, we propose verifiable sufficient conditions in two cases: (i) both the risk mappings and the state vectors are comonotone (a condition automatically satisfied on totally ordered state spaces), and (ii) a one-step look-ahead policy is optimal. We illustrate and verify the results on several standard examples from operations management.



\section*{References}
\bibliographystyle{ieeetran}
\bibliography{demobib}

\end{document}